\documentclass[12pt]{article}
\usepackage{amsmath,amsfonts,amssymb}

\begin{document}
\thispagestyle{empty}

\def\theequation{\arabic{section}.\arabic{equation}}
\def\a{\alpha}
\def\b{\beta}
\def\g{\gamma}
\def\d{\delta}
\def\dd{\rm d}
\def\e{\epsilon}
\def\ve{\varepsilon}
\def\z{\zeta}
\def\teta{\tilde\eta}

\newcommand{\h}{\hspace{0.5cm}}

\begin{titlepage}
\vspace*{1.cm}
\renewcommand{\thefootnote}{\fnsymbol{footnote}}
\begin{center}
{\Large \bf Finite-size effect of $\eta$-deformed $AdS_5\times S^5$ at strong coupling}
\end{center}
\vskip 1.2cm \centerline{\bf Changrim  Ahn }

\vskip 10mm

\centerline{\sl Department of Physics} \centerline{\sl Ewha Womans
University} \centerline{\sl DaeHyun 11-1, Seoul 120-750, S. Korea}
\vspace*{0.6cm} \centerline{\tt ahn@ewha.ac.kr}

\vskip 20mm

\baselineskip 18pt

\begin{center}
{\bf Abstract}
\end{center}
We compute L\"uscher corrections for a giant magnon in the 
$\eta$-deformed $(AdS_5\times S^5)_{\eta}$ using the $su(2|2)_q$-invariant $S$-matrix 
at strong coupling and compare with
the finite-size effect of the corresponding string state, derived previously.
We find that these two results match and confirm that  the $su(2|2)_q$-invariant $S$-matrix is describing 
world-sheet excitations of the $\eta$-deformed background.

\end{titlepage}
\newpage
\baselineskip 18pt

\def\nn{\nonumber}
\def\tr{{\rm tr}\,}
\def\p{\partial}
\newcommand{\non}{\nonumber}
\newcommand{\bea}{\begin{eqnarray}}
\newcommand{\eea}{\end{eqnarray}}
\newcommand{\bde}{{\bf e}}
\renewcommand{\thefootnote}{\fnsymbol{footnote}}
\newcommand{\be}{\begin{eqnarray}}
\newcommand{\ee}{\end{eqnarray}}
\newcommand{\beq}{\begin{equation}}
\newcommand{\eeq}{\end{equation}}

\vskip 0cm

\renewcommand{\thefootnote}{\arabic{footnote}}
\setcounter{footnote}{0}

\setcounter{equation}{0}
\section{Introduction}
$AdS/CFT$ duality \cite{AdS/CFT}, a correspondence between string theories in AdS background with certain supersymmetric 
and conformal Yang-Mills theories on the boundary space-time of the AdS space, 
has been a hot topic for theoretical researches and produced many important 
quantitative results and applications (for overview see \cite{RO}).
In these developments, integrability has played a crucial role on both sides of the correspondence.
Two-dimensional world-sheet actions for the string theory moving in the background are described by
nonlinear sigma models on coset group manifolds which are classically integrable.
Aspects of quantum integrable structure of supersymmetric Yang-Mills theories appear in Bethe ansatz 
equations and related exact integrable machineries which can determine conformal dimensions of the CFTs.
Quantum $S$-matrices of the world-sheet actions provide integrable framework which interpolate from the strong to
weak coupling limits.

An important direction of research is to find new AdS/CFT pairs which show novel integrability structures. 
One such string theory, which has been studied recently, is the type IIB superstring theory in
the $\eta$-deformed targe space $(AdS_5\times S^5)_{\eta}$ for a real parameter $\eta$ \cite{DMV0913}.
The classical integrability of nonlinear sigma model is provided by solutions of the classical Yang-Baxter equation \cite{Klimcik}. (See \cite{KY1,Sfetsos,HRT} for related issues.)
It has been conjectured in \cite{DMV0913} that full quantum $S$-matrix of the deformed sigma model is given by 
the $R$-matrix of the $q$-deformed Hubbard model which has been proposed much earlier in \cite{BeiKor}.
When $q$ is a complex phase, the dressing phase of the $S$-matrix and bound-states have been analyzed in \cite{HHM}.
Scattering amplitudes of bosonic exitations for small values of the world-sheet 
momentum have been computed and shown to agree 
with the $q$-deformed $S$-matrix in the large string tension (strong coupling) limit for real $q$ with explicit relation
with $\eta$ \cite{ABF}.
Based on the exact $S$-matrix, thermodynamic Bethe ansatz equations for ground states and dressing phase for 
real $q$ have been studied in \cite{AdLT}.

A pertinent issue which should be mentioned is that the deformed sigma model is not a fully consistent 
string theory at quantum level.
It has been found that this $\eta$-deformed sigma model does not solve the type IIB supergravity
equations of motion \cite{ABF2015}, but rather a generalization of them \cite{AFHRT}.
This generalized ones allow only scale invariance but not full Weyl invariance at one-loop \cite{WulTse}.
The Weyl invarince can be restored if the deformation is generalized by some modified solutions of the Yang-Baxter equation \cite{BorWul}.
This suggests that one should pay attention to treat the $\eta$-deformed theory at quantum level.

In this letter, we provide another evidence for the $q$-deformed $S$-matrix to describe the string theory on the
$\eta$-deformed geometry.
For this purpose, we consider finite-size effects of a giant magnon state, a classical string configuration
 living on a subspace of the $(AdS_5\times S^5)_{\eta}$ \cite{HM}.
These corrections have been computed for the undeformed $AdS_5\times S^5$ in \cite{AFZ,JL} and for the 
$\gamma$-deformed $AdS_5\times S^5$ in \cite{BF, ABK} from both directions of string solutions and 
world-sheet $S$-matrices.
For the $\eta$-deformed case, this effect has been studied from only string theory side in \cite{AhnBoz}, 
which will be reviewed in sect.2.
Exact $q$-deformed $S$-matrix and related formula will be presented in sect.3.
We present our computation of the L\"uscher corrections for a giant magnon based on $q$-deformed
$S$-matrix in sect.4 along with a conjecture on the deformed dressing phase in sect.5.
In sect.6, we conclude with a short summary and comments.

\section{Finite-size effect of a giant magnon in  $(AdS_5\times S^5)_{\eta}$}
In this section, we give a brief review on computing the energy of a giant magnon using Neumann-Rosochatius ansatz following
\cite{AhnBoz}.
The  giant magnon is defined in the $R_t\times S^3_\eta$ subspace of $(AdS_5\times S^5)_{\eta}$, 
where backgound metric and $B$-field are given by 
\bea
\nn 
&&g_{tt}=-1,\h g_{\phi_1\phi_1}=\sin^2\theta,\h g_{\phi_2\phi_2}=\frac{\cos^2\theta}{1+\tilde{\eta}^2 \sin^2\theta},\\ 
\label{fb} 
&&g_{\theta\theta}=\frac{1}{1+\tilde{\eta}^2 \sin^2\theta},\h b_{\phi_2\theta}=
-\tilde{\eta} \frac{\sin 2\theta}{1+\tilde{\eta}^2 \sin^2\theta}.
\eea
Deformation parameter ${\tilde\eta}$ is related to original parameter $\eta$ by ${\tilde\eta}=2\eta/(1-{\eta}^2)$.

One can solve the giant magnon configuration using an ansatz for the dynamics of the target space coordinates 
\bea\label{A} t(\tau,\sigma)=\kappa \tau,\h \phi_i(\tau,\sigma)=\omega_i \tau+F_i(\xi),
\h \theta(\tau,\sigma)=\theta(\xi),\h \xi=\sigma-v\tau ,\h i=1,2,\eea
where $\tau$ and $\sigma$ are the string world-sheet coordinates and
the Virasoro constraints.
If we restrict further to $S^2$ by setting the isometry angle $\phi_2$ to zero,
conserved charges $E_s,\ J_1$ corrsponding to other isometric coordinates $t,\phi_1$
are given by complete elliptic integrals of first and third kinds ($W=\kappa^2/\omega_1^2$):
\bea
\label{Esf} E_s&=&\frac{2T}{\tilde{\eta}} \frac{(1-v^2)\sqrt{W}}{\sqrt{(\chi_\eta-\chi_m)(\chi_p-\chi_n)}}
\ \mathbf{K}(1-\epsilon),\\
\nn J_1&=&\frac{2T}{\tilde{\eta}\sqrt{(\chi_\eta-\chi_m)(\chi_p-\chi_n)}}
\Bigg[\left(1-v^2W-\chi_\eta\right)\ \mathbf{K}(1-\epsilon)
+(\chi_\eta-\chi_p)\ \mathbf{\Pi}\left(\frac{\chi_p-\chi_m}{\chi_\eta-\chi_m},
1-\epsilon\right)\Bigg],
\eea
where the parameters are satisfying
\be
&&\chi_m=\frac{\chi_\eta \chi_p}{\chi_\eta-(1-\epsilon)\chi_p}\ \epsilon,\qquad \epsilon=\frac{\chi_m(\chi_\eta-\chi_p)}{\chi_p(\chi_\eta-\chi_m)},\nn\\
 &&\frac{(1-\epsilon)\chi_p^2-2\epsilon\chi_p\chi_\eta-\chi_\eta^2}{\chi_\eta-(1-\epsilon)\chi_p}
+3-(1+v^2)W+\frac{1}{\tilde{\eta}^2}=0,\nn\\
&&\chi_p\chi_\eta+\frac{\epsilon\chi_p\chi_\eta(\chi_p+\chi_\eta)}{\chi_\eta-(1-\epsilon)\chi_p}
-\frac{2-(1+v^2)W+\left(3-\left(2+v^2(2-W)\right)W\right)\tilde{\eta}^2}{\tilde{\eta}^2}=0,\nn\\
&&\frac{\epsilon\chi_p^2\chi_\eta^2}{\chi_\eta-(1-\epsilon)\chi_p}
-\frac{(1+\tilde{\eta}^2)(1-W)(1-v^2 W)}{\tilde{\eta}^2}=0.\nn
\ee
The momentum of a giant magnon, which is related to the deficit angle by $\Delta\phi_1=p$, satisfies
\be
p&=&\frac{2v}{\tilde{\eta}\sqrt{\chi_p(\chi_\eta-\chi_m)}}\Bigg\{-v \ \mathbf{K}(1-\epsilon)+\label{mom}\\
&+&\frac{W}{(\chi_\eta-1)(1-\chi_p)}\Bigg[(\chi_\eta-\chi_p)\
\mathbf{\Pi}\left(-\frac{(\chi_\eta-1)(\chi_p-\chi_m)}{(\chi_\eta-\chi_m)(1-\chi_p)},1-\epsilon\right)
-(1-\chi_p)\ \mathbf{K}(1-\epsilon)\Bigg]\Bigg\}.
\nn
\ee
Eqs.(\ref{Esf}) and (\ref{mom}) generate the dispersion relation of a giant magnon at finite $J_1$.

In the limit of $J_1\gg g\gg 1$ one can solve the parameter relations in terms of
small $\epsilon$-expansions to determine the energy and angular momentum 
\bea
\label{fr} 
E_s-J_1&=& \frac{2 g \sqrt{1+\tilde{\eta}^2}}{\tilde{\eta}}\ 
\mbox{arcsinh}\left(\tilde{\eta} \sin\frac{p}{2}\right)\nonumber\\
&-&\frac{8g(1+\tilde{\eta}^2)^{3/2}
\sin^3\frac{p}{2}}{\sqrt{1+\tilde{\eta}^2 \sin^2\frac{p}{2}}} \exp\left(-\frac{J_1}{g}
\sqrt{\frac{1+\tilde{\eta}^2\sin^2\frac{p}{2}}{\left(1+\tilde{\eta}^2\right)\sin^2\frac{p}{2}}}
\right).
\eea
The first term is the energy dispersion relation of a giant magnon in the infinite volume and
the second one is the small finite-size (or finite $J_1$) correction.
In next sections, we are going to reproduce this result from  the $su(2|2)_q$ $S$-matrix. 

\section{$q$-deformed $S$-matrix}

The quantum-deformed $S$-matrix can be written as a graded tensor product of $su(2|2)_q$-invarint matrix as follows:
\be
{\cal S}(p_1,p_2)=S_{su(2)}\ {\bf S}{\hat \otimes} {\dot{\bf S}}.
\ee
The overall scalar factor $S_{su(2)}$ is given by \cite{ABF} \footnote{
Several different candidates have been proposed in \cite{HHM}. We have checked that only this one is consistent with the finite-size correction. }

\be
S_{su(2)}(p_1,p_2)=\frac{1}{\sigma^2(p_1,p_2)}\frac{x_1^++\xi}{x_1^-+\xi}\frac{x_2^-+\xi}{x_2^++\xi}
\frac{x_1^--x_2^+}{x_1^+-x_2^-}
\frac{1-\frac{1}{x_1^-x_2^+}}{1-\frac{1}{x_1^+x_2^-}},
\label{dressing}
\ee
with $q$-deformed dressing phase $\sigma$.
The $su(2|2)_q$-invarint $S$-matrix has $16\times 16$ elements, $S_{ij}^{i'j'}$, $i,j,i',j'=1,\dots,4$.
For L\"uscher correction, the matrix elments we need are
\be
S_{11}^{11}&=& a_1,\quad S_{12}^{12}= \frac{q}{2} a_1+\frac{1}{2} a_2,\quad S_{13}^{13}=S_{14}^{14}=a_5
\label{Selement}\\
a_1&=&1,\quad a_2=-q+\frac{2}{q}\frac{x_1^-(1-x_2^-x_1^+)(x_1^+-x_2^+)}{x_1^+(1-x_2^-x_1^-)(x_1^--x_2^+)},\quad
a_5=\frac{x_1^+-x_2^+}{\sqrt{q}U_1V_1(x_1^--x_2^+)}.
\label{Sexplicit}
\ee
The parameters $x^{\pm}$ satisfy a shortening relation 
\be
\frac{1}{q}\left(x^{+}+\frac{1}{x^{+}}\right)-q\left(x^{-}+\frac{1}{x^{-}}\right)=\left(q-\frac{1}{q}\right)\left(\xi+\frac{1}{\xi}\right),
\label{shorten}
\ee
and related to energy ${\cal E}$ and momentum $p$ by 
\be
V^2=q\frac{x^+}{x^-}\frac{x^-+\xi}{x^++\xi}\equiv q^{{\cal E}},\qquad U^2=\frac{1}{q}\frac{x^++\xi}{x^-+\xi}\equiv e^{ip}.
\label{dispersion}
\ee
The constant $\xi$ is related to the string tension $g$ and deformation parameter $q$ by
\be
\xi=-\frac{i}{2}\frac{g(q-q^{-1})}{\sqrt{1-\frac{g^2}{4}(q-q^{-1})^2}}.
\label{xi}
\ee
It is claimed that the quantum group parameter $q$ is related to ${\tilde\eta}$ by
\be
q=e^{-\nu/g}\quad {\rm with}\quad \nu=\frac{\tilde\eta}{\sqrt{1+{\tilde\eta}^2}}.
\label{param}
\ee
General energy-momentum dispersion relation follows from this
\be
{\cal E}(p)=\frac{2 g}{\nu} \ \mbox{arcsinh}\left(\frac{\xi}{i}\sqrt{\frac{1}{4g^2\cosh^2\frac{\nu}{2g}}+\sin^2\frac{p}{2}}\right).
\label{exenergy}
\ee

At strong coupling  limit $g\gg 1$, Eqs.(\ref{xi}) and (\ref{param}) lead to
\be
\xi=i{\tilde\eta}+{\cal O}(g^{-1}).
\ee
From Eqs.(\ref{shorten}) and (\ref{dispersion}), one can expand the paramters
\be
x^{\pm}(p)=x^{\pm}_0(p)+\frac{1}{g}\ x^{\pm}_1(p)+{\cal O}(g^{-2}),
\label{largexpm}
\ee
where
\be
x^{\pm}_0(p)&=&e^{\pm ip/2}\left(\sqrt{1+{\teta}^2\sin^2\frac{p}{2}}\mp \teta \sin\frac{p}{2}\right),
\label{xpm0}\\
x^{\pm}_1(p)&=&\frac{\left(x^{\pm}_0(p)+i\teta\right)\left(\teta\ x^{\pm}_0(p)-i\right)}{\sqrt{1+{\tilde\eta}^2}\left(x^{-}_0(p)-x^{+}_0(p)\right)}.
\ee
Also the dispersion relation in Eq.(\ref{exenergy}) becomes
\be
{\cal E}_0(p)=\frac{2 g \sqrt{1+\tilde{\eta}^2}}{\tilde{\eta}}\ 
\mbox{arcsinh}\left(\teta \sin\frac{p}{2}\right),
\label{energy}
\ee
which is consistent with that of  giant magnon string state given in the first term of Eq.(\ref{fr}).

\section{L\"uscher corrections}
Leading finite-size corrections in the strong coupling limit are the $\mu$-term L\"uscher corrections which arise from residues 
of $S$-matrix in the contour integrals of the $F$-term formula.
Explicit $\mu$-term L\"uscher formula for one $su(2)$ giant magnon state with $su(2|2)$ index $(1{\dot 1})$ is given by  \cite{Luscher,JL}, 
\beq
\delta E_{\mu}=-i\left(1-\frac{{\cal E}'(p)}{{\cal E}'({\tilde q}_{\star})}
\right)e^{-i{\tilde q}_{\star}J_1}\sum_{j,{\dot j},j',{\dot j'}}
\mathop{{\rm Res}}_{q={\tilde q}}
\left[{\cal S}_{(1\dot 1)(j{\dot j})}
^{(1\dot 1)(j'{\dot j'})}(p,q_{\star}(q))\right],
\label{muterm}
\eeq
where ${\tilde q}$ is the location of $S$-matrix the poles.
The physical giant magnon has momentum $p$ and energy given by (\ref{energy}), while the momentum $q_{\star}$ of the virtual particle satisfies the following on-shell relation
\beq
q^{2}+{\cal E}^{2}(q_{\star})=0.
\label{disper}
\eeq
We also use a short notation ${\tilde q}_{\star}=q_{\star}({\tilde q})$.

We start with locating the poles of the $S$-matrix. 
The overall scalar factor $S_{su(2)}(p,q_{\star})$ in (\ref{dressing}) have both $s$-channel pole at $x^-({\tilde q}_{\star})=x^+(p)$ and
$t$-channel pole at $x^-({\tilde q}_{\star})=1/x^+(p)$.
We have checked that the $t$-channel gives exactly same results as the $s$-channel.
We will present a detailed computation for the $s$-channel here and multiply a factor $2$ at the end.

Substituting $x^+(p)$ for $x^-({\tilde q}_{\star})$ in Eq.(\ref{shorten}), we can compute $x^+({\tilde q}_{\star})$
\beq
x^+({\tilde q}_{\star})=x^{+}_0(p)+\frac{3}{g}\ 
\frac{\left(x^{\pm}_0(p)+i\teta\right)\left(\teta\ x^{\pm}_0(p)-i\right)}{\sqrt{1+{\tilde\eta}^2}\left(x^{-}_0(p)-x^{+}_0(p)\right)}+{\cal O}(g^{-2}).
\label{xpqstar}
\eeq
From Eq.(\ref{dispersion}), we can obtain
\beq
e^{i{\tilde q}_{\star}}=\frac{1}{q}\frac{x^+({\tilde q}_{\star})+\xi}{x^-({\tilde q}_{\star})+\xi}=1+\frac{1}{g}\ 
\frac{\teta\left(x^{+}_0(p)+x^{-}_0(p)\right)-2i}{\sqrt{1+{\tilde\eta}^2}\left(x^{-}_0(p)-x^{+}_0(p)\right)}+{\cal O}(g^{-2}).
\eeq
Using Eq.(\ref{xpm0}), we obtain ${\tilde q}_{\star}$ as follows:
\beq
i{\tilde q}_{\star}=\frac{1}{g}\ \frac{\sqrt{1+{\teta}^2\sin^2\frac{p}{2}}}{\sqrt{1+{\teta}^2}\sin\frac{p}{2}}+{\cal O}(g^{-2}).
\label{qstar}
\eeq
This leads to the exponentially suppressing factor in the L\"uscher formula
\beq
e^{-i{\tilde q}_{\star}J_1}=\exp\left(-\frac{J_1}{g}
\frac{\sqrt{1+\tilde{\eta}^2\sin^2\frac{p}{2}}}{\sqrt{1+\tilde{\eta}^2}\sin\frac{p}{2}}\right),
\label{expon}
\eeq
which matches with the string computations shown in (\ref{fr}).

The next factor to consider in the L\"uscher formula (\ref{muterm}) is the energy dispersion.
Since ${\tilde q}_{\star}\sim {\cal O}(g^{-1})$, one should use exact relation (\ref{exenergy}) instead of (\ref{energy}) before taking the large $g$ limit along with (\ref{qstar}). 
A straightforwad computation yields
\beq
\left(1-\frac{{\cal E}'(p)}{{\cal E}'({\tilde q}_{\star})}
\right)=\frac{\left(1+\tilde{\eta}^2\right)\sin^2\frac{p}{2}}{1+\tilde{\eta}^2\sin^2\frac{p}{2}}.
\label{Epsilon}
\eeq

Now we move on to the residue of the $S$-matrix, which comes from the scalar factor (\ref{dressing})
\be
\mathop{{\rm Res}}_{q={\tilde q}}
\left[S_{su(2)}(p,q_{\star})\right]
=\frac{2e^{3ip/2}\left[1+ie^{ip/2}\teta\left(\sqrt{1+\tilde{\eta}^2\sin^2\frac{p}{2}}-\teta\sin\frac{p}{2}\right)\right]}{g\sqrt{1+\tilde{\eta}^2}\sin\frac{p}{2}
\cdot\sigma^2(p,{\tilde q}_{\star})\cdot {x^{-}}'({\tilde q}_{\star})}.
\ee
The last factor can be computed by a trick used in  \cite{JL}
\be
\left.\frac{dx^{-}(q_{\star})}{dq}\right\vert_{q={\tilde q}}=\frac{dx^{+}(p)/dp}{dq/dp}=
\frac{-ie^{ip/2}\left[1+ie^{ip/2}\teta\left(\sqrt{1+\tilde{\eta}^2\sin^2\frac{p}{2}}-\teta\sin\frac{p}{2}\right)\right]\sin^2\frac{p}{2}}
{\sqrt{1+\tilde{\eta}^2\sin^2\frac{p}{2}}},
\ee
where we used (\ref{disper}) for $dq/dp$.
Combining these, we get 
\be
\mathop{{\rm Res}}_{q={\tilde q}}
\left[S_{su(2)}(p,q_{\star})\right]=
\frac{2ie^{ip}\sqrt{1+\tilde{\eta}^2\sin^2\frac{p}{2}}}{g\sqrt{1+\tilde{\eta}^2}\sin^3\frac{p}{2}\cdot \sigma^2(p,{\tilde q}_{\star})}.
\label{residue}
\ee

The contribution form each matrix element is from Eq.(\ref{Selement}) 
\be
\left(\frac{1+q}{2}\ a_1+\frac{1}{2}a_2+2a_5\right)^2
\ee
and becomes $1$ in the leading order from (\ref{Sexplicit}).

\section{$q$-deformed Dressing phase}

The dressing phase has been proposed first in terms of $q$-deformed Gamma function
for $q$ a complex phase, \cite{HHM}
\be
\sigma^2(p_1,p_2)&=&\exp i\left[\chi(x_1^+,x_2^+)-\chi(x_1^+,x_2^-)-\chi(x_1^-,x_2^+)+\chi(x_1^-,x_2^-)\right],\label{sigma}\\
\chi(x_1,x_2)&=&i\oint_{|z|=1}\frac{dz}{2\pi i} \frac{1}{z-x_1} \oint_{|w|=1} \frac{dw}{2\pi i} \frac{1}{w-x_2} \log
\frac{\Gamma_{q^2}\left[1+\frac{i}{2a}\left(u(z)-u(w)\right)\right]}{\Gamma_{q^2}\left[1-\frac{i}{2a}\left(u(z)-u(w)\right)\right]},\label{chi}
\ee
where $a=\nu/g$ for $g\gg 1$ and $u(z)$ is defined by
\be
X(z,w)\equiv u(z)-u(w)=\frac{i}{2\nu}\log\left(\frac{z+\frac{1}{z}+\xi+\frac{1}{\xi}}{w+\frac{1}{w}+\xi+\frac{1}{\xi}}\right).\label{uu}
\ee
An integral representation for $\Gamma_{q^2}$ given in \cite{HHM} can be analytically continued for real $q$ to get 
strong coupling limit \cite{ABF}
\be
\log\frac{\Gamma_{q^2}\left[1+gX\right]}{\Gamma_{q^2}\left[1-gX\right]}&\approx&
g\Biggl\{2X(\log g-1)+X\log(-X^2)\nonumber\\
&-&i\frac{2\pi}{\nu}\left[\psi^{-2}\left(1-\frac{i\nu X}{\pi}\right)-\psi^{-2}\left(1+\frac{i\nu X}{\pi}\right)\right]\Biggr\},
\label{gGammaq}
\ee
where $\psi^{-2}$ is the poly-gamma function.
The integrals over two unit circles in (\ref{chi}) may develop a branch cut for $\nu\ge 1/2$ 
but can be handled with proper care as pointed out in \cite{AdLT}.

For computing $\sigma(p,{\tilde q}_{\star})$ at strong coupling, we combine the $\chi$ functions with arguments
$x^{\pm}(p),\ x^{\pm}({\tilde q}_{\star})$ given in Eqs.(\ref{largexpm}) and (\ref{xpqstar}) to get 
\be
&&\log\sigma^2({\tilde q}_{\star},p)=\oint_{|z|=1}\frac{dz}{2\pi i} \oint_{|w|=1}\frac{dw}{2\pi i}
\left[\frac{2i\nu x_0^+\left(x_0^++x_0^-+\xi+\frac{1}{\xi}\right)}{g(z-x_0^+)(z-x_0^-)
(w-x_0^+)^2}\right]\times\nonumber\\
&\times&g\left\{X\log\frac{g^2}{e^2}+X\log(-X^2)-\frac{2\pi i}{\nu}\left[\psi^{-2}\left(1-\frac{i\nu X}{\pi}\right)-\psi^{-2}\left(1+\frac{i\nu X}{\pi}\right)\right]
\right\},
\label{integralform}
\ee
with a short notation $x^{\pm}_0=x^{\pm}_0(p)$ given in (\ref{xpm0}).
Due to complicated branch cuts appearing in the contour integrals, we could not evaluate this integral analytically.
However, we have found numerically that the integration depends on $\teta$ very insensitively within available accuracy.
This leads to our conjecture that the dressing phase with given arguments in the strong coupling limit is
\be
\sigma^{2}({\tilde q}_{\star},p)=-2 g^2\ e^{-ip}\ \sin^4\frac{p}{2},
\label{dressinglarge}
\ee
which is the result for the undeformed case, computed from the AFS phase \cite{AFS} in  \cite{JL}.

Combining (\ref{expon}), (\ref{Epsilon}), (\ref{residue}) and (\ref{dressinglarge}) along with a factor $-i$ in (\ref{muterm}) and $2$ for the $t$-channel 
contribution, we get the final $\mu$-term L\"uscher correction 
\be
\delta E_{\mu}=-\frac{8g(1+\tilde{\eta}^2)^{1/2}
\sin^3\frac{p}{2}}{\sqrt{1+\tilde{\eta}^2 \sin^2\frac{p}{2}}} \exp\left(-\frac{J_1}{g}
\sqrt{\frac{1+\tilde{\eta}^2\sin^2\frac{p}{2}}{\left(1+\tilde{\eta}^2\right)\sin^2\frac{p}{2}}}
\right).
\label{final}
\ee

\section{Conclusion}
Compared with finite-size giant magnon computation (\ref{fr}), the strong coupling L\"uscher correction match quite well
except $1+\teta^2$ in the overall factor.
We think this factor should be modified in the string theory computation. 
Apart from this minor discrepancy, 
both coefficient and exponent of the exponential factor show correct dependence on the momentum and deformation parameter.
Our check is valid for the $su(2)$ sector with generic value of $p$ and 
supports that the $q$-deformed $S$-matrix should describe the string theory in the $\eta$-deformed AdS background.
It will be interesting to further elaborate the $q$-deformed dressing phase to check (\ref{dressinglarge}) both numerically and
analytically.
Another interesting but less studied domain is the weak coupling limit of the $S$-matrix, which could be related to 
certain $q$-deformed spin-chain.

\section*{Acknowledgements}
We thank S. van Tongeren for sharing useful information on the dressing phase.
This work was supported by the National Research Foundation of Korea (NRF) grant
(NRF-2016R1D1A1B02007258).

\end{document}